\documentclass[twocolumn,preprintnumbers,amsmath,amssymb]{revtex4}
\usepackage{graphicx}% Include figure files
\usepackage{dcolumn}% Align table columns on decimal point
\usepackage{bm}% bold math

\def\etal{{\it et\ al.}}

\newcommand{\lsim}
 {\ \raise.35ex\hbox{$<$}\kern-0.75em\lower.5ex\hbox{$\sim$}\ }
\newcommand{\gsim}
 {\ \raise.35ex\hbox{$>$}\kern-0.75em\lower.5ex\hbox{$\sim$}\ }
\def\journal #1#2#3#4{#1 {\bf #2} (#4) #3}
\def\PR{Phys.\ Rev.}
\def\PRB{Phys.\ Rev.\ B}
\def\PRL{Phys.\ Rev.\ Lett.}

\def\JPParis{J.~Phys.~(Paris)}

\def\JPCS{J.\ Phys.\ Chem.\ Solids}

\def\JPSJ{J.\ Phys.\ Soc.\ Jpn.}

\def\PTP{Prog.\ Theor.\ Phys.}

\def\RPP{Rep.~Prog.~Phys.}
\begin{document}

\title{Variational Monte Carlo Studies of Pairing Symmetry for the 
$t$-$J$ Model \\on a Triangular Lattice}% Force line breaks with \\

\author{Tsutomu \textsc{Watanabe}$^{1,2}$, 
Hisatoshi \textsc{Yokoyama}$^{3}$, Yukio \textsc{Tanaka}$^{1,2}$, 
Jun-ichiro \textsc{Inoue}$^{1}$ and \\
Masao \textsc{Ogata}$^{4}$}%
\affiliation{%
$^{1}$Department of Applied Physics, Nagoya University,  Nagoya 464-8603 \\
$^{2}$CREST Japan Science and Technology Corporation (JST) \\
$^{3}$Department of Physics, Tohoku University, Sendai 980-8578 \\
$^{4}$Department of Physics, University of Tokyo, Bunkyo-ku, Tokyo 113-0033 \\
}%

\date{\today}% It is always \today, today,

\begin{abstract}
As a model of a novel superconductor Na$_x$CoO$_2$$\cdot$$y$H$_2$O, 
a single-band $t$-$J$ model on a triangular lattice is studied, 
using a variational Monte Carlo method. 
We calculate the energies of various superconducting (SC) states, 
changing the doping rate $\delta$ and sign of $t$ for small $J/|t|$.
Symmetries of $s$, $d$, and $d$+$id$ ($p$+$ip$ and $f$) waves are 
taken up as candidates for singlet (triplet) pairing. 
In addition, the possibility of Nagaoka ferromagnetism and 
inhomogeneous phases is considered. 
It is revealed that, among the SC states, the $d$+$id$ wave always 
has the lowest energy, which result supports previous mean-field 
studies. 
There is no possibility of triplet pairing, although the $f$-wave 
state becomes stable against a normal state in a special case 
($\delta=0.5$ and $t<0$). 
For $t<0$, the complete ferromagnetic state is dominant in a 
wide range of $\delta$ and $J/|t|$, which covers the realistic 
parameter region of superconductivity. 
\end{abstract}

\maketitle

\section{\label{sec:level1}Introduction}

The recent discovery of superconductivity \cite{Takada} in 
Na$_{0.35}$CoO$_2\cdot$1.3H$_2$O has aroused an active interest of 
solid state physicists, since this compound is regarded as a strongly 
correlated superconductor on a triangular lattice. 
For such magnetically frustrated lattices, novel superconducting (SC) 
features are expected, as compared to the celebrated cuprates on 
bipartite lattices. 
In particular, it is primarily important to determine the symmetry of 
its pairing potential. 
So far, there have been a number of experimental efforts trying to 
fix the pairing symmetry, 
e.g. measurements of nuclear spin lattice relaxation rate $T_1$
\cite{Y.Kobayashi,Ishida,Fujimoto}, 
Knight Shift \cite{Waki,Y.Kobayashi} by NMR, and by 
$\mu$SR \cite{Higemoto,Uemura}. 
At present, however, the results of these researches are not necessarily 
consistent to each another; the pairing symmetry has not yet been established. 
\par

As for theory, although some studies regard multiband effects as important,
\cite{Koshibae,Mochizuki,Yata} many treat this issue with reduced 
single-band models on a triangular lattice. 
The single-band studies are roughly classified into two categories, 
namely ones applying weak-coupling techniques to Hubbard-type models, 
like perturbation expansions \cite{Ikeda}, a random-phase approximation 
\cite{Y.Tanaka} and a fluctuation exchange approximation (FLEX) \cite{Kuroki}, 
and ones starting with $t$-$J$-type models in the strong-coupling 
limit.\cite{Kumar,Baskaran,Wang,Ogata,SNS} 
Although the weak-coupling studies took mutually different approaches 
for different models, they arrived at a conclusion that $f$-wave 
spin-triplet symmetries are competitive with or even dominant over 
$d$-wave spin-singlet symmetries. 
On the other hand for $t$-$J$-type models, it is unanimously shown 
that $d_{x^2-y^2}$+$id_{xy}$-wave symmetries are widely stable 
against other symmetries as well as metallic states, using slave-boson 
mean-field approximations\cite{Baskaran, Wang, Kumar}, which barely 
respect the local constraint of no double occupation, and a Gutzwiller 
approximation\cite{Ogata}, which takes account of the constraint 
on an average. 
Thus, the mean-field results of the strong-coupling theories disagree 
with the results of the weak-coupling ones, in contrast with 
the case of a square lattice, where the conclusion of the $d_{x^2-y^2}$ 
wave has been drawn in both theories. 
\par 

In this paper, we study the $t$-$J$ model on a triangular lattice, 
using a variational Monte Carlo (VMC) method, which exactly treats 
the local constraint of no double occupation. 
Our main purposes are 
(i) to determine the pairing symmetry of superconductivity within the 
$t$-$J$ model with this more accurate method, and  
(ii) to check the possibility of spin-triplet superconductivity in 
this strong-coupling model. 
Besides the SC states, (iii) we consider the stability of Nagaoka 
ferromagnetism\cite{Nagaoka,MH,Koretsune,Ferro} 
and inhomogeneous phases.
Based on these calculations, we argue the propriety of modeling 
for cobaltates. 
\par

The organization of this paper is as follows: 
In \S2, the formulation used is described. 
Sections 3 and 4 are assigned to the results and discussions on 
the stability of the singlet and triplet SC states, 
respectively. 
In \S5, we treat ferromagnetism and phase separation. 
In \S6, we recapitulate our results and address the relevance 
to cobaltates. 
A part of the present results has been reported.\cite{SNS}
\par

%%%%%%%%%%%%%%%%%%%%%%%%%%%%%%%%%%%%%%%%%%%%%%%%%%%%%%%%%%%%%%%%%%%%%%%%%
\section{\label{sec:formulation} Formulation}
\begin{figure}
\begin{center}
\includegraphics[width=8cm,height=4.5cm]{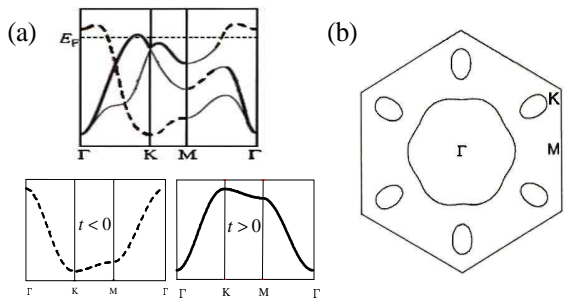}
\end{center}
\caption{
(a) Upper panel: Schematic plot of the band structure for 
Na$_{0.5}$CoO$_2$, following an LDA calculation\cite{Singh}. 
Lower panels: Dispersion of single-band tight binding models on 
a triangular lattice for $t<0$ and $t>0$.  
(b) Hole Fermi surfaces of the CoO planes in Na$_{0.5}$CoO$_2$, due 
to the same LDA.}
\label{fig:band} 
\end{figure}

%%%%%%%%%%%%%%%%%%%%%%%%%%%%%%%%%%%%%%%%%%%%%%%%%%%%%%%%%%%%%%%%%%%%%%%%%
Generally, the 3$d$ orbitals are split into $t_{2g}$ and $e_g$ 
orbitals by an octahedral crystal field. 
In this material, the $t_{2g}$ orbitals of Co are further split into 
the levels of $a_{1g}$ $\left[=(d_{xy}+d_{yz}+d_{zx})/{\sqrt{3}}\right]$ 
and $e_g'$ $\left[=\left(d_{xy}+e^{\pm i\frac{2\pi}{3}}d_{yz}
+e^{\pm i\frac{4\pi}{3}}d_{zx}\right)/\sqrt{3}\right]$ 
\cite{note:Eg},  
by a shift of oxygen vertical to the CoO$_2$ layers. 
Due to an LDA calculation for Na$_{0.5}$CoO$_2$\cite{Singh}, 
the bands near the Fermi surface [Fig.~\ref{fig:band}(b)] are 
involved in a Co $t_{2g}$ manifold [Fig.~\ref{fig:band}(a)].
Since Na$_{0.35}$CoO$_2\cdot$1.3H$_2$O has the same structure as 
Na$_{0.5}$CoO$_2$, except for the intercalation of water 
molecules and the sodium concentration, one may assume 
the conduction bands for Na$_{0.35}$CoO$_2\cdot$1.3H$_2$O are 
similarly formed from the $t_{2g}$ manifold. 
The band which forms the large Fermi surface around the $\Gamma$ point 
(Fig.~\ref{fig:band}(b)) has dominant $a_{1g}$ character, whereas 
the band which forms the small pocket Fermi surfaces on the $\Gamma$-K 
lines have mixed $a_{1g}$ and $e'_g$ character. 
Thus, we call the band indicated by dashed (thick solid) lines in 
Fig.~\ref{fig:band}(a) the $a_{1g}$ ($e'_g$) band. 
Many previous studies have assumed that Na$_{0.35}$CoO$_2\cdot$1.3H$_2$O 
is basically described by single-band models of the $a_{1g}$ or 
$e'_g$ band, as mentioned. 
\par

%***************************************************************************
%  Fig.1
%***************************************************************************
%******************************************************************************

\begin{figure}
\begin{center}
\includegraphics[width=4.5cm,height=2.8cm]{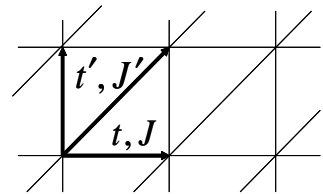}
\end{center}
\caption{
Lattice structure of the model used in this study, with 
its parameters.
}
\label{fig:model}
\end{figure}

Along this line, we consider a single-band $t$-$J$ model on an extended 
square lattice, 
\begin{eqnarray}
{\cal H}={\cal H}_t+{\cal H}_J=-\sum_{(i,j)\sigma}t_{ij}
{\cal P}_{\rm G}(c_{i\sigma}^\dag c_{j\sigma}+{\rm H.c.})
{\cal P}_{\rm G} \nonumber\\
+\sum_{(i,j)}J_{ij}\left({{\bf S}_i\cdot{\bf S}_j-\frac{1}{4}n_in_j}
\right)
\label{eq:model}, 
\end{eqnarray}
where 
${\cal P}_{\rm G}=\prod_i{(1-n_{i\uparrow }n_{i\downarrow})}$. 
The sum of site pairs $(i,j)$ is restricted as follows: $t_{ij}=t$ and 
$J_{ij}=J (>0)$ for the nearest-neighbor pairs, $t_{ij}=t'$ and 
$J_{ij}=J' (>0)$ for the next-nearest-neighbor pairs only in one diagonal 
direction, and $t_{ij}=J_{ij}=0$ otherwise (see Fig.~\ref{fig:model}).
Except for the case of $t<0$ and $\delta\sim 0.5$ (\S4), 
where the degeneracy at $\varepsilon=\varepsilon_{\rm F}$ is serious, 
we put $t'=t$ and $J'=J$. 
Here, $\delta$ indicates the hole concentration: $\delta=|1-n|$ with 
$n=N_{\rm e}/N_{\rm s}$ ($N_{\rm e}$: electron number, 
$N_{\rm s}$: site number) .
This parameter setting makes no difference with the isotropic triangular 
lattice. 
\par 

Let us make a remark on the sign of $t$. 
In the lower panels of Fig.~\ref{fig:band}(a), the tight-binding bands 
of the present model, 
\begin{equation}
\varepsilon_{\bf k}=-2t\left(\cos k_x+\cos k_y\right)
                    -2t'\cos\left(k_x+k_y\right), 
\label{eq:dispersion}
\end{equation}
with $t'=t$ are depicted for $t>0$ and $t<0$. 
Compared with the LDA band structure (upper panel of Fig.~\ref{fig:band}(a)), 
it seems appropriate that the band eq.~(\ref{eq:dispersion}) with 
$t<0$ [$t>0$] should be connected to the $a_{1g}$ (dashed lines) 
[$e'_g$ (bold lines)] band. 
In Na$_x$CoO$_2$$\cdot$$y$H$_2$O, the electron density is larger than 1, 
i.e. electron-doped case. 
To treat electron-doped cases in the framework of $t$-$J$-type models, 
namely in the electron density of $0\le n\le 1$ without double 
occupation, we take advantage of a particle-hole transformation. 
Thereby, an electron-doped case with $t < 0$, 
which corresponds to the $a_{1g}$ band 
(lower left panel of Fig.~\ref{fig:band}(a)), is mapped to a hole-doped 
case with $t > 0$ in $t$-$J$ model of eq.~(\ref{eq:model}). 
We use this sign convention throughout this paper. In fact, 
the $t<0$ band, which corresponds to the $e'_g$ band, 
has a van Hove singularity at $\delta=0.5$, 
as the $e'_g$ band does near the K points.
So far, the majority of theoretical studies\cite{A.Tanaka,Baskaran,Wang, 
Ogata,Ikeda,Y.Tanaka} have considered that the large Fermi 
surface around the $\Gamma$ point mainly contributes, whereas 
some papers have regarded the isolated pocket Fermi surfaces 
or the van Hove singularities as important for superconductivity.
\cite{Kumar,Kuroki,Mochizuki}
In this paper, we consider both cases, $t<0$ and $t>0$. 
\par

Since the cobalt atoms in the present compound are connected 
through edge-shared oxygen cages, the effective value of $J/|t|$ is 
expected to be fairly small. 
Accordingly, we principally discuss the region of small $J/|t|$ 
in this paper.
\par

%************************************************************************
%  Fig.2
%************************************************************************
%************************************************************************

We move on to the method used. 
To consider strongly correlated systems, more accurate treatment 
of the local constraint is necessary beyond the mean field level. 
For this purpose, a variational Monte Carlo (VMC) method 
\cite{McMillan,Ceperley,Yokoyama2} is appropriate.
In this method, variational expectation values of many-body wave 
functions are numerically estimated with a Monte Carlo procedure, 
in which the local constraint, e.g. ${\cal P}_{\rm G}$ in 
eq.~(\ref{eq:model}), is exactly satisfied. 
Actually, it has been shown that the VMC method is effective for 
a variety of analyses of the $t$-$J$-type models on square lattices 
without (or with small) frustration \cite{YO,Randeria}. 
At least, we believe this method is reliable to discuss the mutual 
stability among various pairing symmetries. 
In this paper, we apply this method to the model eq.~(\ref{eq:model}) 
to study the stability of various SC states. 
\par

For $t$-$J$-type models, a Gutzwiller-type variational wave function 
is a good starting point,\cite{GJR,YO,Randeria} unlike Hubbard-type 
models.\cite{YS3}
For the normal state, we use the original Gutzwiller wave function 
$\Psi_{\rm n}={\cal P}_{\rm G}\Phi_{\rm F}$ ($\Phi_{\rm F}$: Fermi 
sea).\cite{Gutzwiller} 
A SC state with a fixed electron number\cite{Lhuillier} is given as, 
\begin{eqnarray}
\left|\Psi_{\rm sc}\right\rangle=
{\cal P}_{\rm G}|{\rm BCS}\rangle={\cal P}_{\rm G}
\left({\sum\limits_{\bf k}{\varphi _{\bf k}
c_{{\bf k}\uparrow}^\dag c_{-{\bf k}\downarrow}^\dag}}\right)
^{\frac{{N_{\rm e}}}{2}}\left|0\right\rangle,
\label{eq:wavefunc}
\end{eqnarray}
with 
$$
\varphi_{\bf k}=\frac{u_{\bf k}}{v_{\bf k}}=
\frac{\Delta_{\bf k}}{{\varepsilon_{\bf k}+
\sqrt{\varepsilon_{\bf k}^2+\Delta_{\bf k}^2}}}.
$$ 
Anisotropy of the pairing symmetry is introduced in $\Delta_{\bf k}$.
We consider $s$, extended-$s$ (two kinds), $d$ and $d$+$id$ waves as 
candidates for singlet pairing, and $p$+$ip$ and $f$ waves for triplet 
pairing. 
They are summarized in Table I.
Symmetries of $s$, $p$+$ip$ and $d$+$id$ waves have no 
line node in $\Delta_{\bf k}$, whereas $d$ and $f$ waves have line nodes. 
The time-reversal symmetry is broken for $d$+$id$ and $p$+$ip$ waves. 
In the present calculation, $\Delta$ is the only variational parameter, 
which is connected to the SC gap, in particular for a high-doping 
region, but is not the gap itself in contrast with weak-coupling theories. 
Except for half filling, we may consider $\Psi_{\rm sc}$ with finite 
$\Delta$ is a SC state, while at half filling $\Psi_{\rm sc}$ 
is an insulating RVB state.\cite{Anderson} 
We fix $\mu$ at $\mu_{0}$, the chemical potential for the noninteracting 
system, because $\mu$ dependence of $E_{\rm tot}$ is small, and $\mu$ 
does not seem an essential variable in a fixed-$N_{\rm e}$ formulation. 
\par

%************************************************************************
%  Table I
%\\\\\\\\\\\\\\\\\\\\\\\\\\\\\\\\\\\\\\\\\\\\\\\\\\\\\\\\\\\\\\\\\\\\\\\\
\begin{table}
\caption{\label{tab:table1} 
$\Delta_{\bf k}$ for various pairing symmetries.}
\begin{tabular}{c|c} \hline
 Symmetry & $\Delta _{\bf k}$ \\
\hline
 $s$ & $\Delta$ \\
 ext. $s$ (1) & $\Delta(\cos k_x+\cos k_y)$ \\
 ext. $s$ (2) & $\Delta[\cos k_x+\cos(k_x+k_y)+\cos k_y]$ \\
 $d$ & $\Delta (\cos k_x-\cos k_y)$ \\
 $d$+$id$ & $\Delta\left[{\cos k_x+e^{i\frac{{2\pi}}{3}}\cos(k_x+k_y)
+e^{i\frac{{4\pi}}{3}}\cos k_y}\right]$ \\
 $p$+$ip$ & $\Delta\left[{\sin k_x+e^{i\frac{{2\pi}}{3}}\sin(k_x+k_y)
+e^{i\frac{{4\pi}}{3}}\sin k_y}\right]$ \\
 $f$ & $\Delta \left[ {\sin k_x-\sin(k_x+k_y)+\sin k_y}\right]$ \\
\hline
\end{tabular}
\end{table}
%\\\\\\\\\\\\\\\\\\\\\\\\\\\\\\\\\\\\\\\\\\\\\\\\\\\\\\\\\\\\\\\\\\\\\\\\
Since the number of variational parameters is at most one in this 
study, an usual VMC procedure \cite{Yokoyama2,Lhuillier} sufficiently 
works. 
To obtain an accurate energy difference (condensation energy), we have 
taken a sufficient number of samples to reduce the statistical errors, 
and kept the sampling interval long enough to ensure statistical 
independence of the samples. 
Here, we collect samples as many as $10^5$-$10^7$, which yield 
the error in energy of approximately $10^{-4}\left|t\right|$.
We use the systems of $N_s=L\times L$ ($L$=4-18) with the 
periodic-antiperiodic boundary conditions, and the electron densities 
satisfying the closed shell condition. 
\par

%==========================================================
\section{\label{sec:singlet} Stability of singlet pairings}
%==========================================================
In this section, we discuss the symmetries of singlet pairing, 
$s$, extended-$s$, $d$ and $d$+$id$ waves. 
\par

\begin{figure}
\begin{center}
\includegraphics[width=7.5cm,height=10cm]{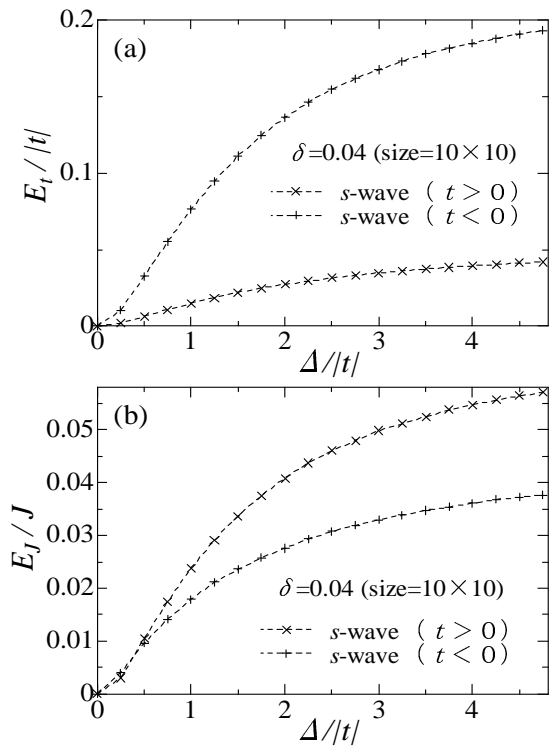}
\end{center}
\caption{(a) Hopping energy and (b) exchange energy of the homogeneous 
$s$-wave state as a function of $\Delta/|t|$ for $\delta=0.04$. 
The two cases, $t>0$ and $t<0$, are simultaneously shown. 
We take the origin of each energy as the value for $\Delta=0$, namely 
the normal state, $E_t(0)$ and $E_J(0)$; similar metrics are used in 
some figures following. 
The system used has $10 \times 10$ sites.
}
\label{fig:swave}
\end{figure}

To begin with, we consider $s$-type pairings. 
Figure \ref{fig:swave} shows the hopping energy 
$E_t=\langle{\cal H}_t\rangle$ and exchange energy 
$E_J=\langle{\cal H}_J\rangle$ for the homogeneous $s$-wave symmetry, 
as a function of $\Delta$. 
For both $t<0$ and $t>0$, each of $E_t/|t|$ and $E_J/J$ is a monotonically 
increasing function of $\Delta$, indicating the energy minimum is 
at $\Delta=0$ for any value of $J/t$; the $s$-wave state is not 
stable against the normal state. 
We also confirmed for various values of $\delta$ that this situation 
does not change. 
This result is natural for the homogeneous $s$ wave, which favors 
on-site pairing. 
As for the two kinds of extended-$s$ waves, types (1) and (2), although 
$E_t/|t|$ is a rapidly increasing function of $\Delta$, $E_J/J$ has 
a shallow minimum at a finite value of $\Delta$. 
This means these symmetries become stable for finite values of $J/t$, 
unlike the homogeneous $s$ wave. 
We confirmed, however, that the energy minimum is at $\Delta=0$ 
for a small value of $J/|t|$.\cite{note:ext.s}
We thus conclude that there is no possibility of the $s$ and 
extended-$s$ waves realized in the regime of our concern. 
We will not refer to these states, henceforth. 
\par

%***********************************************************************
%     Fig.3  Etot of s wave
%***********************************************************************
%***********************************************************************

\begin{figure}
\begin{center}
\includegraphics[width=7.5cm,height=10cm]{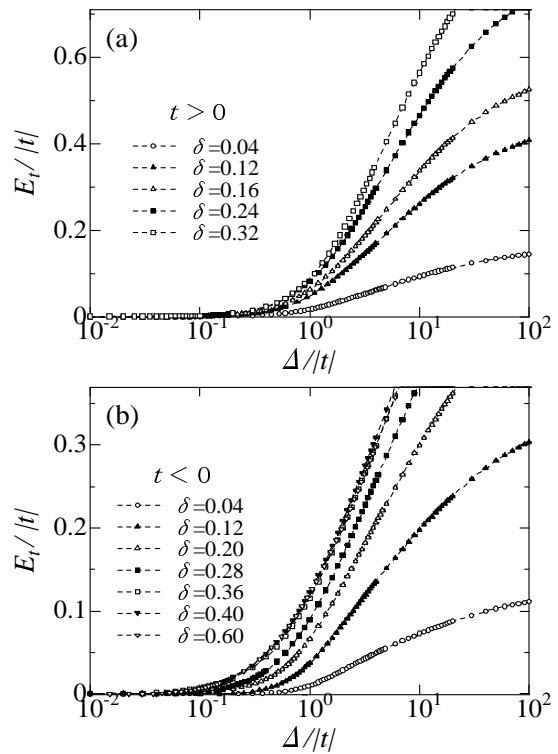}
\end{center}
\caption{Hopping energy of the $d$+$id$-wave symmetry for several values 
of the hole density $\delta$. 
(a) $t>0$ and (b) $t<0$. 
The system size is $10 \times 10$. 
Statistical errors are smaller than the symbol sizes. 
 }
\label{fig:didet}
\end{figure}

\begin{figure}
\begin{center}
\includegraphics[width=7.5cm,height=10cm]{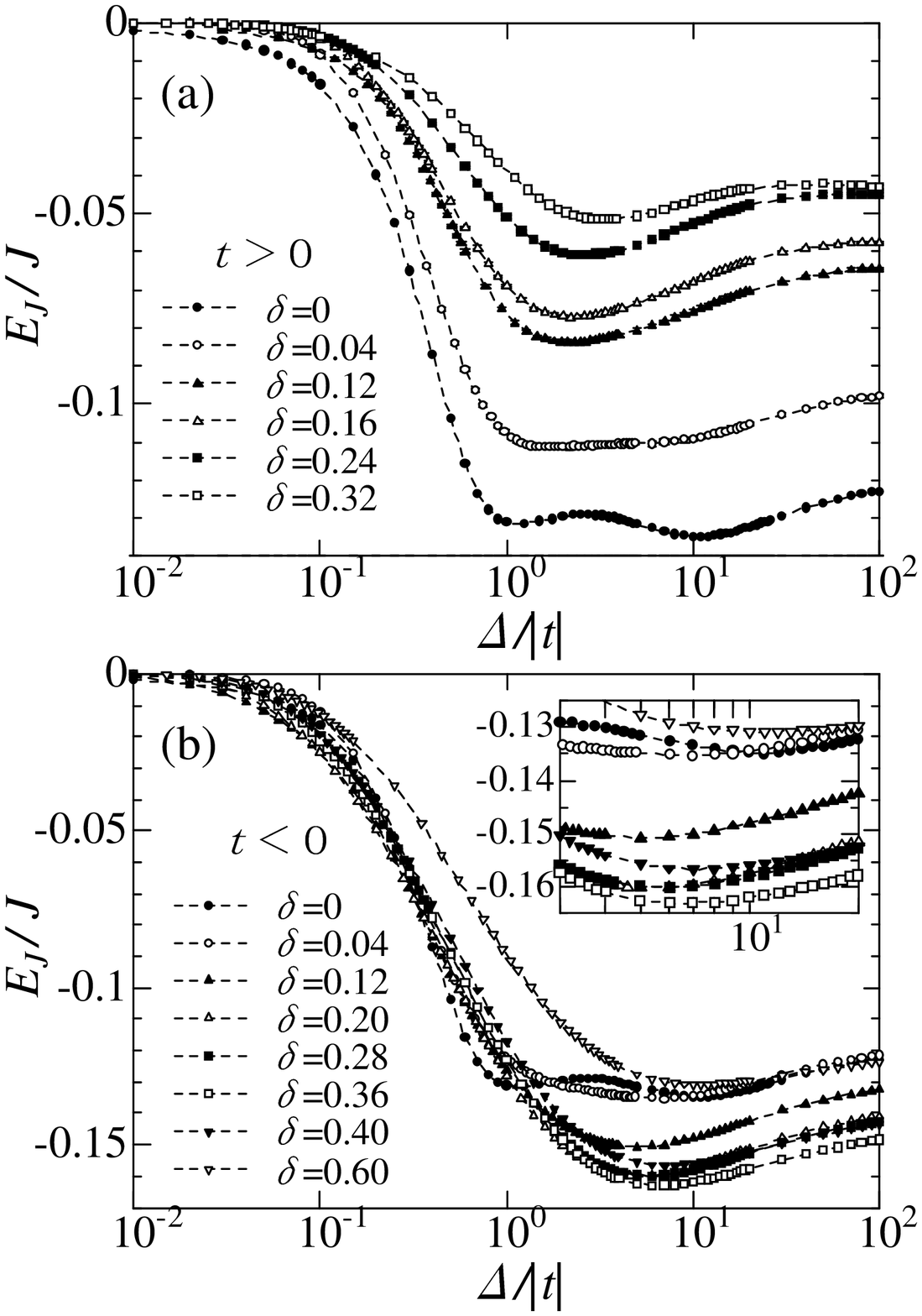}
\end{center}
\caption{Exchange energy of the $d$+$id$-wave symmetry for several values 
of the hole density $\delta$. 
(a) $t>0$ and (b) $t<0$. 
The inset in (b) is the magnification of the minimum area.
The system is the same as Fig.\ref{fig:didet}. 
 }
\label{fig:dideJ}
\end{figure}

\begin{figure}
\begin{center}
\includegraphics[width=7.5cm,height=10cm]{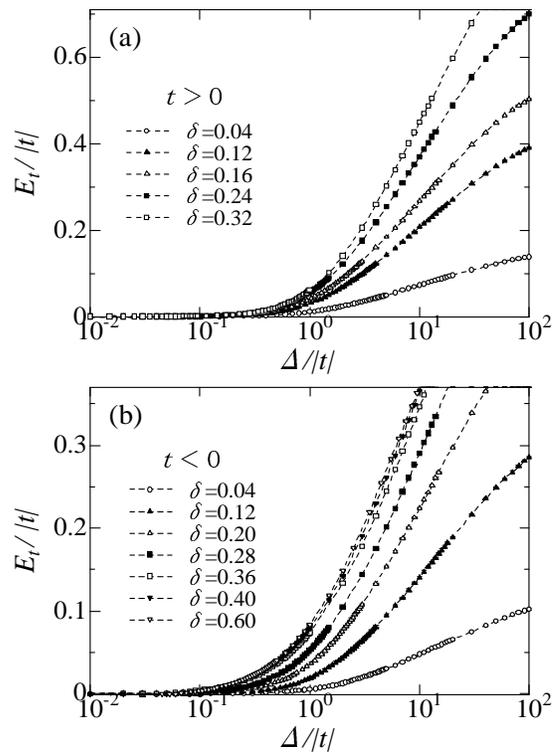}
\end{center}
\caption{Hopping energy of the $d$-wave symmetry for several values 
of the hole density $\delta$. 
(a) $t>0$ and (b) $t<0$. 
The system size is $10 \times 10$. 
 }
\label{fig:det}
\end{figure}

\begin{figure}
\begin{center}
\includegraphics[width=7.5cm,height=10cm]{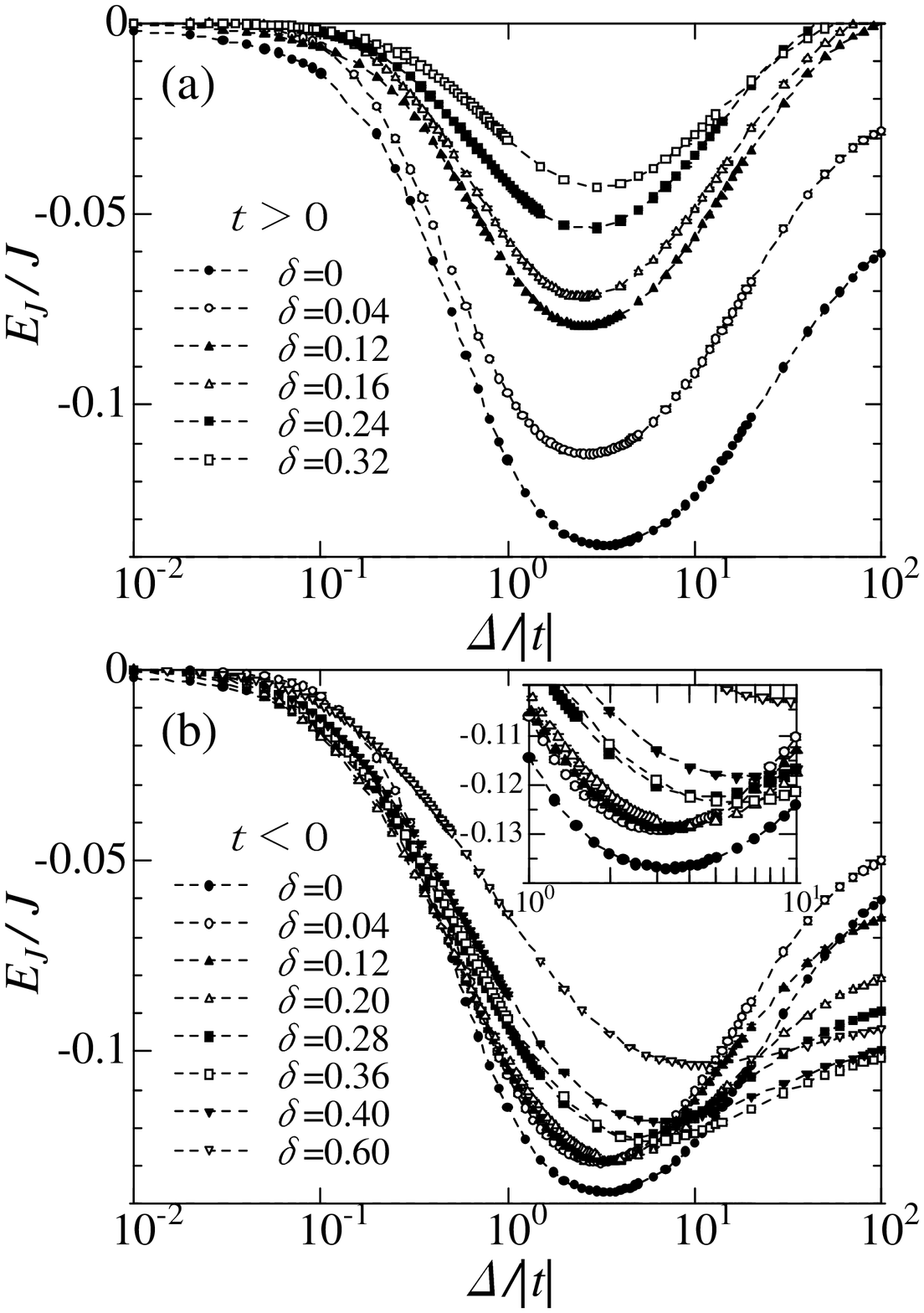}
\end{center}
\caption{Exchange energy of the $d$-wave symmetry for several values 
of the hole density $\delta$. 
(a) $t>0$ and (b) $t<0$. 
The inset in (b) is the magnification of the minimum area.
The system is the same as Fig.\ref{fig:det}. 
 }
\label{fig:deJ}
\end{figure}

\begin{figure}
\begin{center}
\includegraphics[width=7.5cm,height=10.5cm]{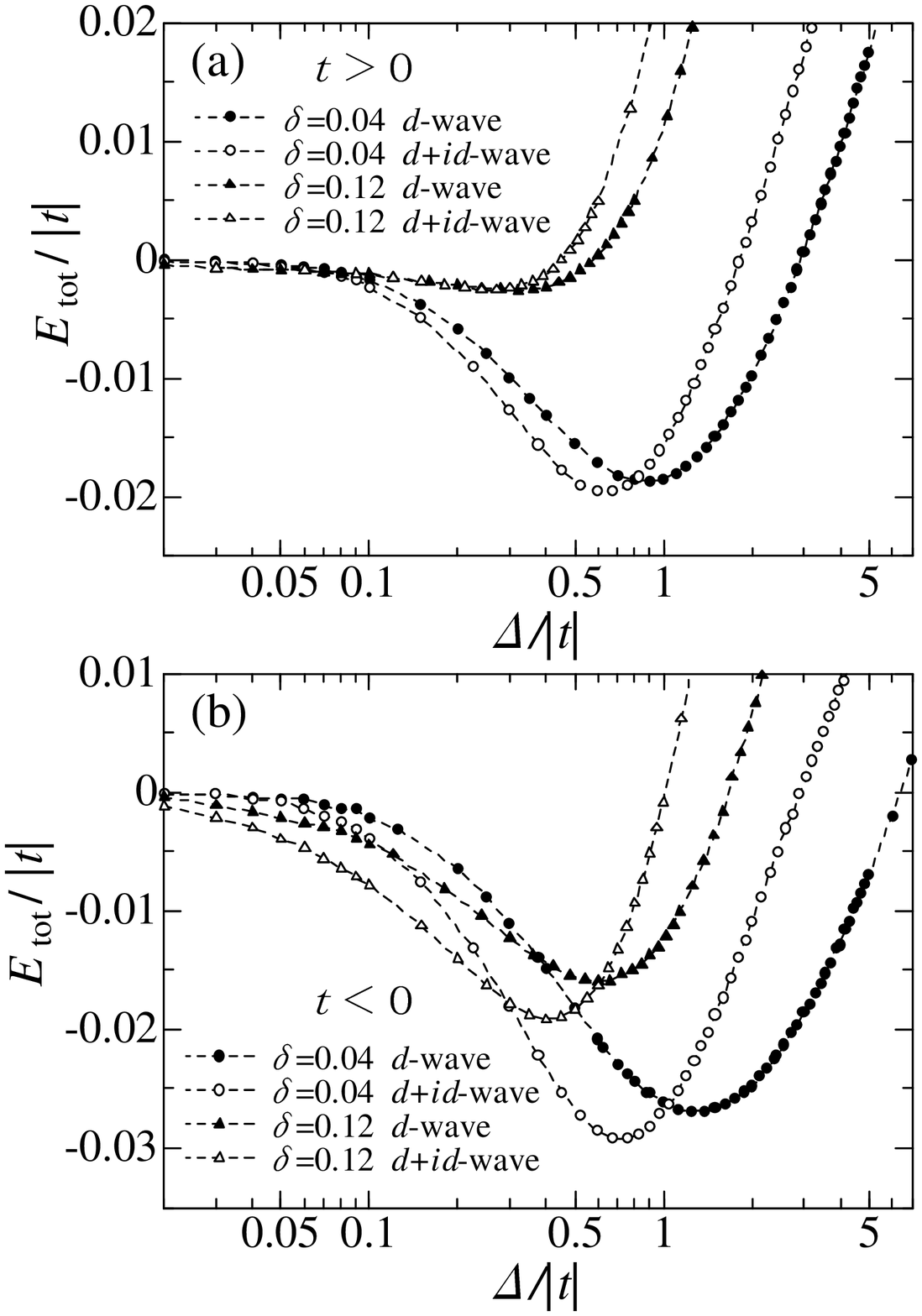}
\end{center}
\caption{Total energy of the $d$- (solid symbols) and $d$+$id$-wave 
(open symbols) states for $\delta=0.04$ and 0.12. 
The system is of $10 \times 10$, and $J/|t|=0.3$. 
(a) $t>0$ and (b) $t<0$. 
}
\label{fig:didetot}
\end{figure}

Next, we turn to the $d$- and $d$+$id$-wave pairings. 
Figures \ref{fig:didet} and \ref{fig:dideJ} show $E_t/|t|$ and $E_J/J$, 
respectively, of the $d$+$id$ wave as a function of $\Delta/|t|$. 
In this case, $E_t/|t|$ is again a monotonically increasing 
function of $\Delta$, whereas $E_J/J$ conspicuously decreases and 
has a minimum at considerably large values of $\Delta$. 
Thereby, the $d$+$id$ state becomes stable even for considerably small 
value of $J/|t|$ (e.g. $J/t\gsim 0.05$ for $\delta=0.04$ and $t>0$). 
Similar behaviors of $E_t$ and $E_J$ can be seen for the $d$-wave state, 
as seen in Figs.~\ref{fig:det} and \ref{fig:deJ}. 
From these figures, we find that the $d$+$id$ and $d$ waves have similar 
energies.
To see actual energy reduction, we plot in Fig.~\ref{fig:didetot} 
the total energy $E_{\rm tot}$ ($=E_t+E_J$) for $J/|t|=0.3$ and 
$\delta=0.04$ and 0.12, as an example. 
For both $d$ and $d$+$id$ waves, $E_{\rm tot}$ has a clear minimum 
at large values of $\Delta$ for both signs of $t$; the SC states of 
$d$ and $d$+$id$ waves are appreciably stable near half filling.
For $\delta\sim 0$, the minimum of the $d$+$id$ wave is slightly 
lower than that of the $d$ wave.\cite{half-filling}
\par

%***********************************************************************
% Fig.4  E_t of d+id waves
%***********************************************************************
%***********************************************************************
% Fig. 5  E_J of d+id waves
%***********************************************************************
%*****************************************************************************
% Fig. 6  E_t of d waves
%*****************************************************************************
%*****************************************************************************
% Fig. 7  E_J of d waves
%*****************************************************************************
%*****************************************************************************
% Fig. 8  Etot of d and d+id waves
%*****************************************************************************
\begin{figure}
\begin{center}
\includegraphics[width=7.5cm,height=10.5cm]{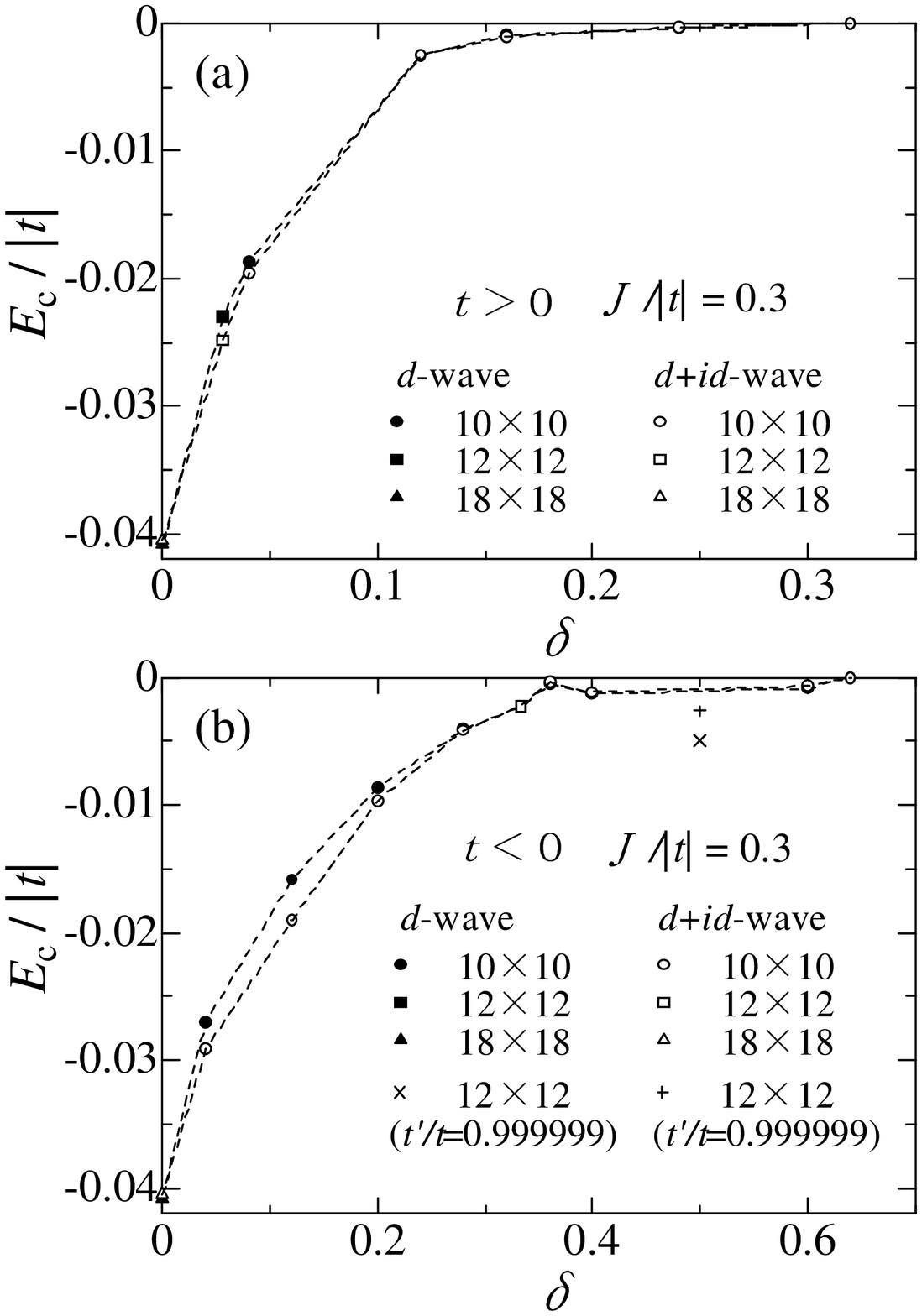}
\end{center}
\caption{
Condensation energy of the $d$- and $d$+$id$-wave states as a function 
of doping rate $\delta$ for $J/|t|=0.3$. 
(a) $t<0$ and (b) $t>0$. 
In (b), the data for $\delta=0.5$ (the van Hove singularity point) is 
obtained under a different condition; for details, see \S\ref{sec:triplet}. 
Note that the range of abscissa is twice larger for $t<0$. 
The Data for $L=10$, 12 and 18 are simultaneously plotted; system-size 
dependence is not severe.}
\label{fig:dec} 
\end{figure}

%*****************************************************************************
%
To consider the $\delta$ dependence of energy, we define a condensation 
energy, $E_{\rm c}$, as the difference of $E_{\rm tot}$ between the 
minimized superconducting state and the normal state, 
\begin{equation} 
E_{\rm c}=E(\Delta_{\rm min})-E(0). 
\end{equation} 
For $E_{\rm c}<0$, the superconducting state is stable against 
the normal state. 
Plotted in Fig.~\ref{fig:dec} is $E_{\rm c}$ for the $d$ and $d$+$id$ 
waves; $|E_{\rm c}|$ is a rapidly decreasing function of $\delta$. 
Near half filling, the $d$+$id$-wave state is more stable than the 
$d$-wave state for both signs of $t$, probably because the symmetry 
of $\Delta_k$ of the former is more favorable to the band symmetry, 
which is not far from isotropic. 
The difference of $E_{\rm c}$, however, again becomes indistinguishable, 
as $\delta$ further increases, and $|E_{\rm c}|$ itself becomes small. 
\par

Note that $E_{\rm c}$ of the $d$ and $d$+$id$ waves remains finite 
up to $\delta=0.24$ for $t>0$, and $\delta=0.60$ for $t<0$. 
Namely, the range of $\delta$ where the $d$+$id$-wave state is 
dominant is much wider for $t<0$ than for $t>0$. 
This behavior agrees with the Gutzwiller approximation\cite{Ogata}
even quantitatively. 
This difference is understood as follows: For $t<0$, the density of 
states $\rho(\varepsilon)$ increases as $\delta$ increases, 
and has a van Hove singularity at $\delta=0.5$, while, for $t>0$, 
$\rho(\varepsilon)$ is a monotonically decreasing function 
of $\delta$ until the band bottom. 
This difference yields different behaviors particularly in $E_J$, 
as seen in Figs.~\ref{fig:dideJ} and \ref{fig:deJ}. 
For $t>0$, the minimum in $E_J$ smoothly increases as $\delta$ increases 
for both $d$ and $d$+$id$ waves. 
On the other hand, for $t<0$, the minimum in $E_J$ once decreases 
as $\delta$ increase, and has the lowest value at $\delta\sim 0.36$ 
for the $d$+$id$ wave [Fig.~\ref{fig:dideJ}(b)]; for the $d$ wave, 
the minimum in $E_J$ is stationary for $0.04\lsim\delta\lsim 0.20$ 
[Fig.~\ref{fig:deJ}(b)]. 
For both waves, the decrease in $E_J$ is still conspicuous for 
$\delta$ as large as 0.6. 
For this reason, the stability of the SC is enhanced particularly 
near the van Hove singularity point; we will return to this point 
in \S\ref{sec:triplet}.
\par 

%*****************************************************************************
%   Fig.9  Ec of d and d+id waves vs. delta
%*****************************************************************************
%*****************************************************************************

\begin{figure}
\begin{center}
\includegraphics[width=7.5cm,height=5cm]{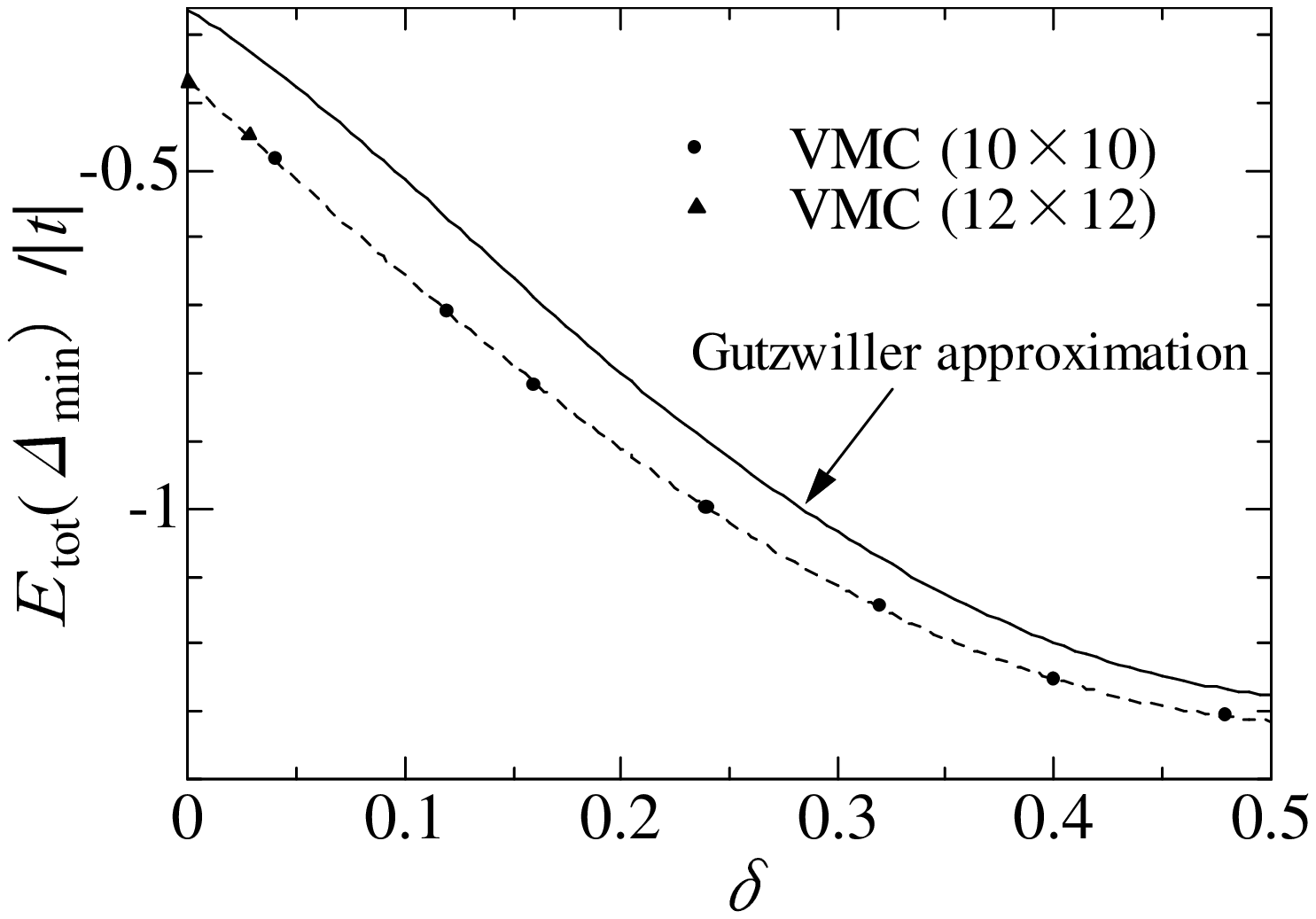}
\end{center}
\caption{
Comparison of $E_{\rm tot}(\Delta_{\rm min})$ of the $d$+$id$-wave 
pairing between the present VMC and a Gutzwiller 
approximation\cite{Ogata} for $t<0$ and $J/|t|=0.3$.}
\label{fig:comparison} 
\end{figure}

The above VMC results that the $d$+$id$ wave is dominant qualitatively 
support the conclusions derived from mean-field 
approximations\cite{Kumar,Baskaran,Wang} and a Gutzwiller 
approximation.\cite{Ogata} 
For comparison, we plot in Fig.\ref{fig:comparison} 
$E_{\rm tot}(\Delta_{\rm min})$ of the $d$+$id$-wave state by the present 
calculations and that of the Gutzwiller approximation. 
Although $E_{\rm tot}(\Delta_{\rm min})$ by VMC is somewhat lower than that 
by Gutzwiller approximation, the two curves exhibit similar doping 
dependence. 
\par

%*****************************************************************************
%   Fig.10  Comparison btw. GA and VMC
%*****************************************************************************
%*****************************************************************************

%*****************************************************************************
\section{\label{sec:triplet} Possibility of triplet pairings}
%****************************************************************************
Ferromagnetic fluctuation, favored by triplet superconductivity, 
has been predicted by an LDA calculation,\cite{SinghTri} and actually 
observed in an NMR-$T_1$ measurement for a superconducting 
sample.\cite{Ishida}
In addition, many theories in the weak-coupling region have made 
conclusions of probable triplet superconductivity. 
In view of these suggestions, we study the stability of triplet 
symmetries, $p$+$ip$ and $f$ waves, in this section. 
\par

For triplet symmetries, the same VMC procedure is carried out 
in the $S^z=0$ sector. 
Having calculated energies for various parameter values, it is found 
that $E_J$ is always a monotonically increasing function of $\Delta$. 
This is because the exchange term ${\cal H}_J/J$ with $J>0$ in 
eq.~(\ref{eq:model}) is obviously unfavorable to a triplet state. 
Moreover, $E_t/|t|$ is also a monotonically increasing functions of 
$\Delta$, except for the case of $\delta\sim 0.5$ and $t<0$. 
Namely, the triplet superconducting states are always unstable against 
the normal state, excluding this special condition. 
In the following, therefore, we concentrate on this special case, 
$t<0$ and $\delta=0.5$, at which the Fermi energy is situated just 
at the van Hove singularity point. 
\par

At this point, severe level degeneracy arises due to the divergence 
of $\rho(\varepsilon_{\rm F})$. 
As a result, for finite systems, the electron densities where the 
closed-shell condition is satisfied disappear near $\delta=0.5$. 
For example, this range is $0.4<\delta<0.6$ for $L=10$, as seen 
in Fig.\ref{fig:dec}(b). 
To avoid this difficulty, we slightly change the value of $t'/t$ 
from 1 to e.g. 0.999999, which lifts the degeneracy and make it possible 
to study $\delta=0.5$ without breaking the closed-shell condition.
\cite{note:degeneracy}  
\par

%*****************************************************************************
%   Fig.11  E_t and E_J for triplets
%****************************************************************************
\begin{figure}
\begin{center}
\includegraphics[width=7.5cm,height=10.5cm]{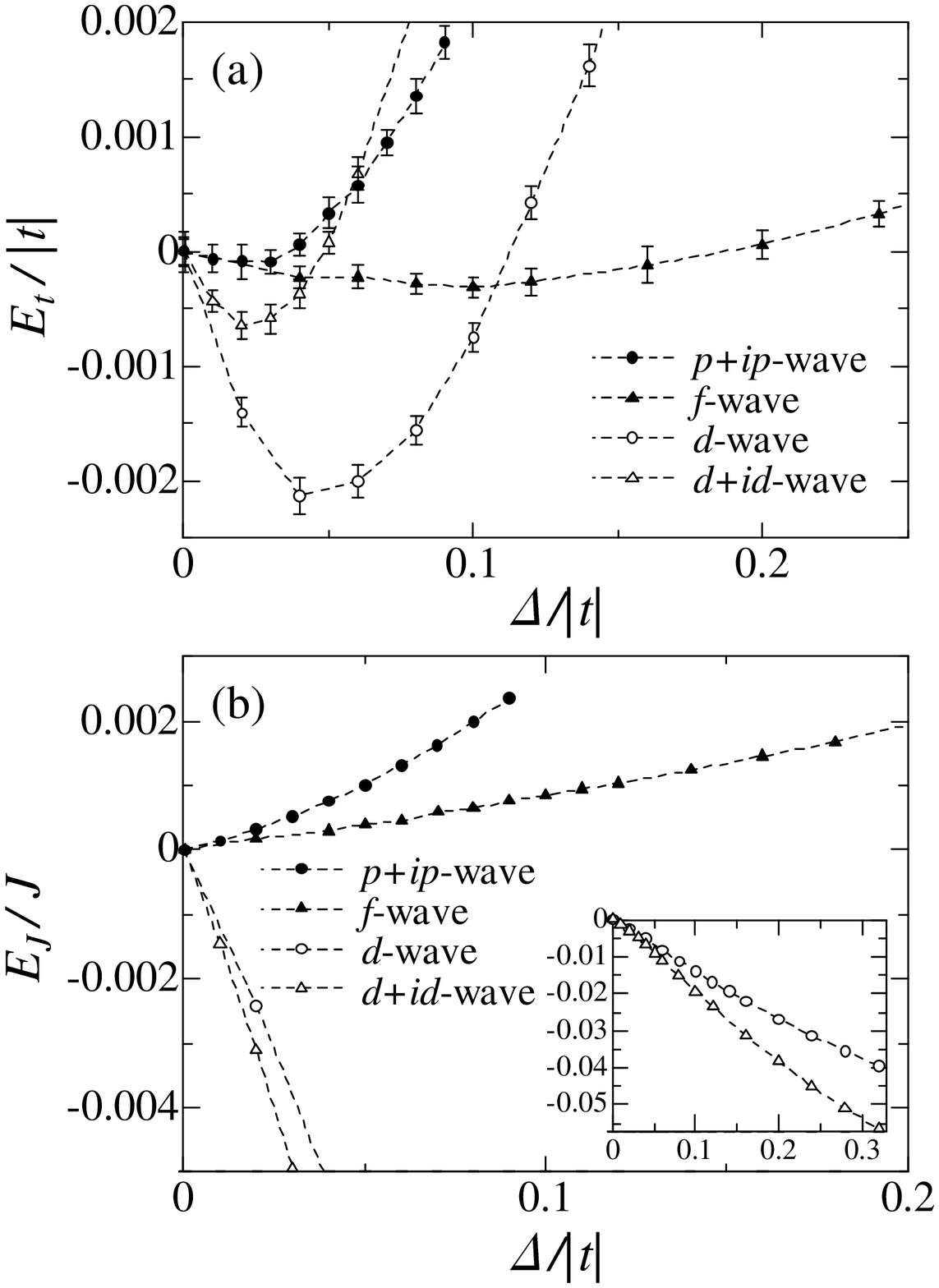}
\end{center}
\caption{Comparison of (a) hopping energy and (b) exchange energy among 
the $p$+$ip$, $f$, $d$+$id$ and $d$ waves for $t'/t=0.999999$ 
($t<0$) and $\delta=0.5$ ($L=12$). 
Here, we assume $J'/J=(t'/t)^2$. 
The inset in (b) shows the behavior of the $d$ and $d$+$id$ waves in 
a wider range.
}
\label{fig:triplet}
\end{figure}
%*****************************************************************************

In Fig.~\ref{fig:triplet}(b), $E_J$ is shown for four symmetries. 
As mentioned above, ${\cal H}_J$ works disadvantageously to the triplet 
pairing symmetries; $E_J$ for the $p$+$ip$ and $f$ waves is again 
a monotonically increasing function of $\Delta$, in contrast with 
the behavior of the singlet pairing symmetries. 
Thus, the triplet SC states are never stabilized when $J/|t|$ increases. 
In Fig.~\ref{fig:triplet}(a), $E_t$ is depicted for the same symmetries. 
Note that $E_t$ for the $f$ and $p$+$ip$ waves (solid symbols) has 
a shallow minimum at a finite value of $\Delta$. 
From this result, we find that the $f$-wave state becomes stable against 
the normal state ($\Delta=0$) for small values of $J/|t|$ 
($J/|t|\lsim 0.18$). 
Compared with the singlet pairing symmetries, however, the decrease 
in $E_t$ for the triplet pairings is much smaller than that for 
the $d$+$id$ and $d$ waves. 
Moreover, the singlet SC states gain energies in $E_J$ for finite 
values of $J/t$. 
We add the data of the $d$ and $d$+$id$ waves at this density 
($J/|t|=0.3$) to Fig.~\ref{fig:dec}(b); we find that the van Hove 
singularity appreciably contributes to superconductivity. 
Nonetheless, we will see that even this $d$ wave is defeated by 
ferromagnetism for small $J/|t|$. 
\par

To summarize, triplet superconductivity is not realized 
in the triangular $t$-$J$ model. 
\par

%*****************************************************************************
\section{\label{sec:ferro} Ferromagnetism and phase separation}
%****************************************************************************
In this section, we discuss ferromagnetic states and inhomogeneous 
phases, which are inseparable from the $t$-$J$ model.
\par

Let us begin with the Nagaoka ferromagnetism.\cite{Nagaoka} 
Since Nagaoka's theorem (in the limit of $J/|t|$, $\delta\rightarrow 0$) 
holds for $t<0$ in a triangular lattice, a finite range of 
ferromagnetism is expected near this limit. 
Actually, it was shown using a variation scheme for the Hubbard model 
on a triangular lattice with $t<0$ that a Nagaoka itinerant 
ferromagnetism is widely stable.\cite{MH} 
For the $t$-$J$ model with $t<0$, recent studies using a high-temperature 
expansion\cite{Koretsune} showed the Nagaoka region to be 
$0<\delta\lsim 0.85$ and $J/t\lsim 0.5$, which is quite wider than 
that for a square lattice (possibly a partial ferromagnetism), namely, 
$0<\delta\lsim0.4$ and $J/t\lsim 0.1$.\cite{Ferro,YO} 
\par

\begin{figure}
\begin{center}
\includegraphics[width=7.5cm,height=10cm]{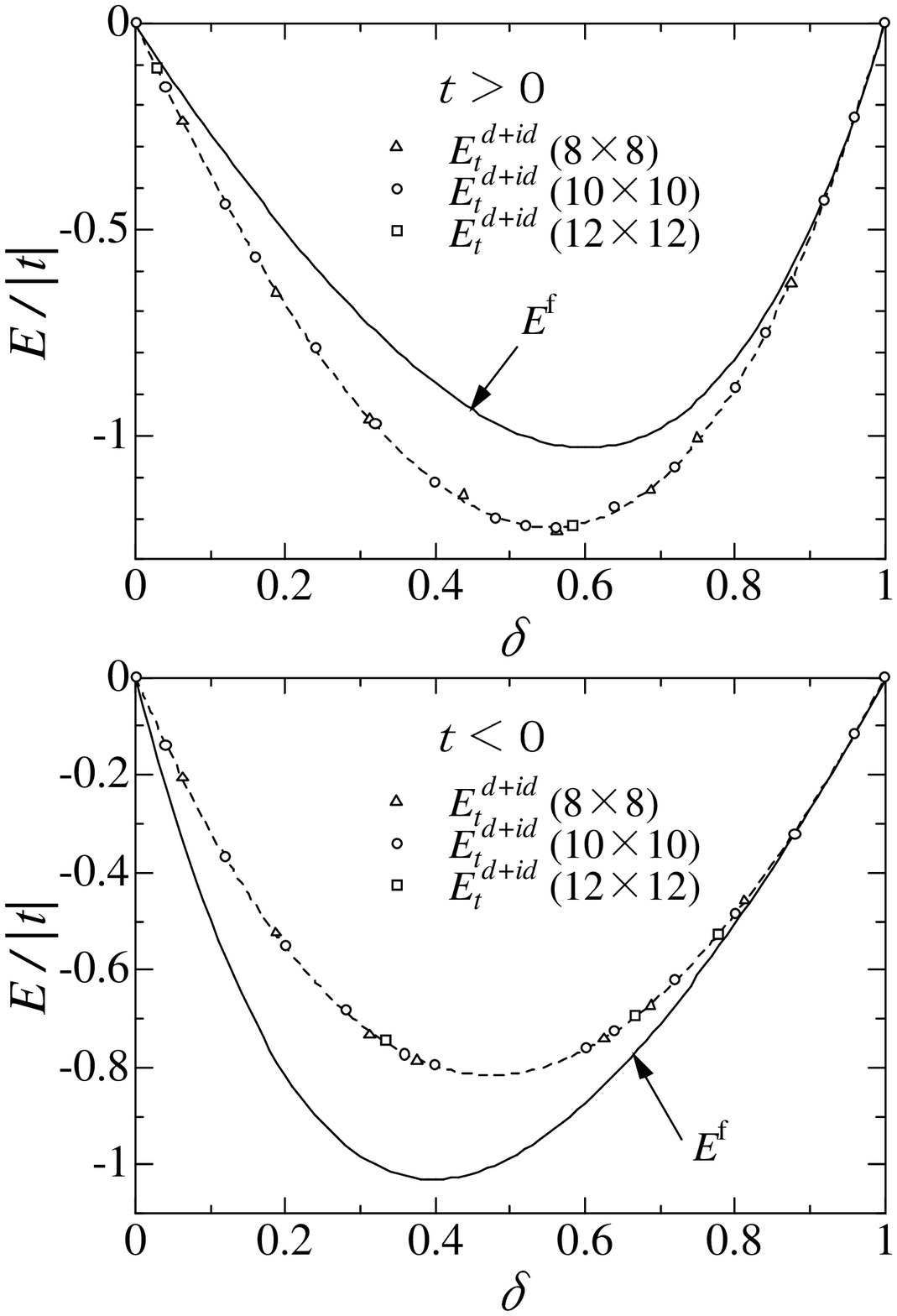}
\end{center}
\caption{Comparison of the transfer energy between the optimized VMC 
values ($J/t=0$) of the $d$+$id$-wave SC state and those of 
the complete ferromagnetic state ($E^{\rm f}$), estimated analytically 
through a spinless Fermion model. 
(a) $t>0$, and (b) $t<0$.
}
\label{fig:ferro} 
\end{figure}

With these in mind, we look at our VMC results in the following. 
In Fig.~\ref{fig:ferro}, we show $E_t$ for the complete ferromagnetism 
(Nagaoka state), $E^{\rm f}$, and the $d$+$id$-wave SC state, $E_t^{d+id}$. 
For $t>0$, $E^{\rm f}$ is always higher than $E_t^{d+id}$; thus the 
complete ferromagnetism never takes place, because 
$\langle{\cal H}_J\rangle=0$ for the Nagaoka state, while $E^{d+id}$ 
further decreases due to the contribution of ${\cal H}_J$. 
Inversely, for $t<0$, $E^{\rm f}$ is appreciably smaller than 
$E_t^{d+id}$ for most values of $\delta$, thereby there exist a region 
for small $J/|t|$ where the Nagaoka state is more stable than the 
$d$+$id$-wave state.
\par

%*****************************************************************************
%   Fig.12  E_t between Nagaoka ferro and d+id wave
%*****************************************************************************
%*****************************************************************************

\begin{figure}
\begin{center}
\includegraphics[width=7.5cm,height=11cm]{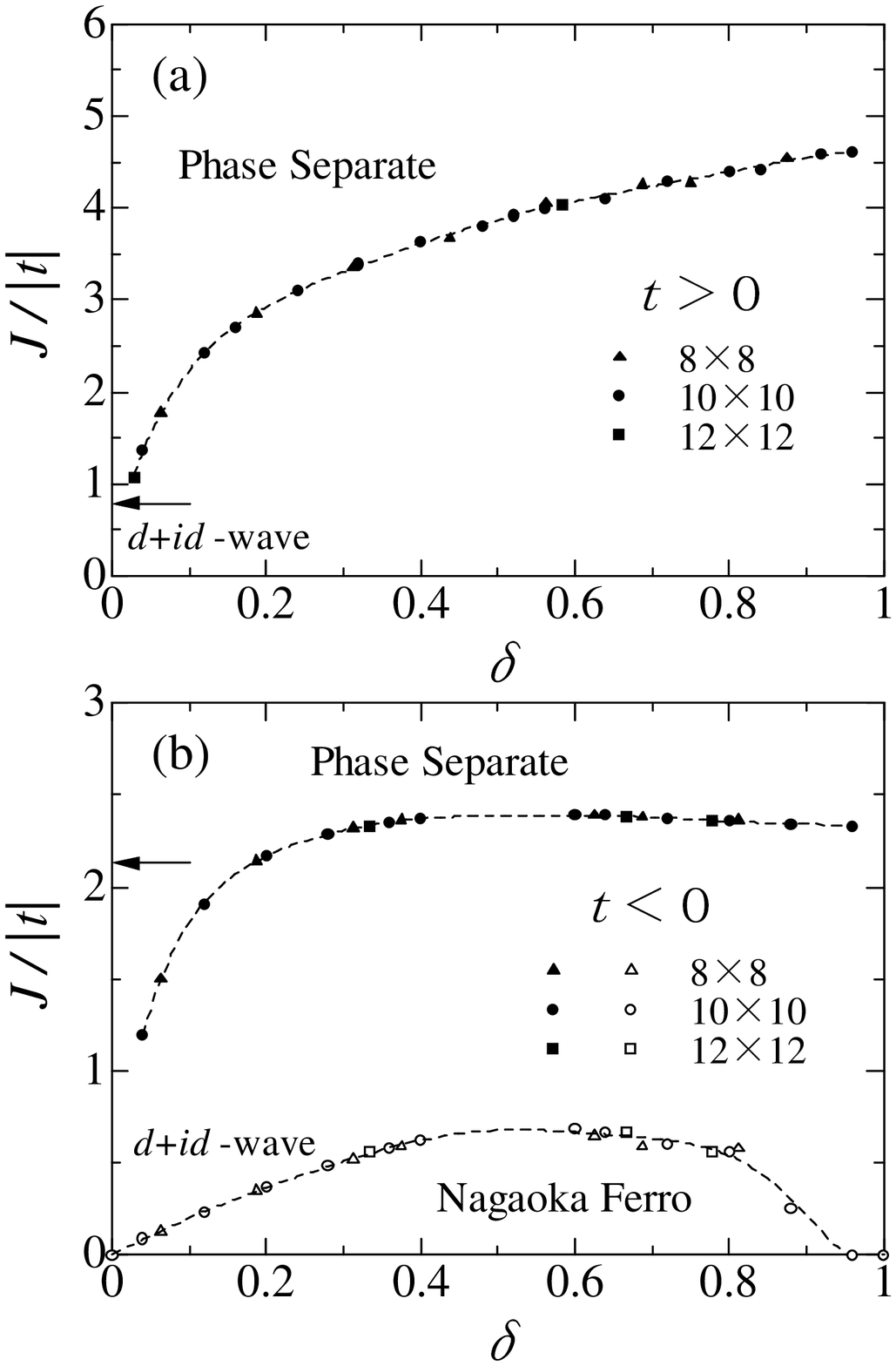}
\end{center}
\caption{Phase diagram in the $\delta$-$J/|t|$ space constructed from 
the present calculations, for (a) $t>0$ and (b) $t<0$. 
The arrows on the vertical axis indicate the critical values of 
the phase separation estimated by the way (ii). 
Note that the system-size dependence is insignificant. 
}
\label{fig:phased}
\end{figure}

Comparing $E^{\rm f}$ and $E^{d+id}$ for finite values of $J/t$, 
we determined the phase boundary, as shown in Fig.~\ref{fig:phased}(b). 
The range of ferromagnetism is very broad, namely 
$\delta\le 0.96$ and $J/|t|\lsim 0.7$, which is quantitatively consistent 
with a recent result of high-temperature expansion,\cite{Koretsune} 
and is markedly wider than that for the square lattice estimated 
through the same VMC method.\cite{Ferro,YO} 
Consequently, for $t<0$, it is unlikely that superconductivity is realized 
in the region of small $J/|t|$ and intermediate $\delta$, even if we 
consider that the ferromagnetic region somewhat shrinks by improving 
the trial function of superconducting states. 
\par

%*****************************************************************************
%   Fig.13  phase diagram
%*****************************************************************************
%*****************************************************************************

Now, we turn to the phase separation in the triangular lattice. 
Since the exchange term in eq.~(\ref{eq:model}) works as a kind of 
attractive potential, the $t$-$J$ model necessarily brings about 
a phase separation for large values of $J/|t|$. 
For the one-dimensional lattice, the boundary between homogeneous and 
inhomogeneous phases extends in the direction of large $J/t$ as $\delta$ 
decreases \cite{Ogata1d}, whereas for the square lattice, the boundary 
approaches the origin [$(J/t,\delta)=(0,0)$] \cite{Emery,YO}. 
\par 

Here, we estimate the boundary in two distinct ways. 
(i) We suppose that the inhomogeneous state phase separates 
into the two hole densities, $\delta=0$ (half filling) and $\delta=1$ 
(empty), and the energy of this state is estimated as 
$E^{\rm PS}=(1-\delta)E^{\rm AF}$, where we adopt, as $E^{\rm AF}$, 
a precise value obtained by a quantum Monte Carlo method for the 
Heisenberg model,\cite{Capriotti} in which the antiferromagnetic (AF) 
state with the $120^\circ$ structure is realized. 
The boundary is determined by comparing $E^{\rm PS}$ and $E^{d+id}$. 
Note that the reliability of this estimation increases as $\delta$ 
increases, but the boundary necessarily converges to $J/|t|=0$ 
as $\delta\rightarrow 0$. \cite{note:hf}
(ii) Instead, we estimate the boundary near half filling, using 
the condition of intrinsic stability, 
$\partial^2 E/\partial \delta^2>0$, for the $d$+$id$-wave state. 
Since we actually substitute finite differences for 
$\partial^2 E/\partial \delta^2$, the critical value of $J/|t|$ 
at half filling for $L\rightarrow\infty$ is likely to be smaller. 
\par

The results of (i) and (ii) are summarized in Fig.~\ref{fig:phased}. 
For $t<0$, the critical value $J_{\rm c}/|t|$ is relatively small 
and barely depends on $\delta$ for large $\delta$. 
For $t>0$, $J_{\rm c}/|t|$ is fairly large for large $\delta$, and 
gradually decreases as $\delta$ decreases. 
For both signs of $t$, $J_{\rm c}/|t|$ abruptly decreases as $\delta$ 
approaches half filling; this tendency is analogous to the square 
lattice case. 
Although the possibility of phase separation still remains for a 
realistic value of $J/|t|$ near half filling, this is not 
the case with intermediate and large values of $\delta$. 
\par

%****************************************************************************
\section{\label{sec:summary} Summary and discussions}
%****************************************************************************
In this paper, we have studied the single-band $t$-$J$ model 
on a triangular lattice, based on variational Monte Carlo (VMC) 
calculations, with Na$_{0.35}$CoO$_2\cdot$1.3H$_2$O in mind. 
The present study, which treats the local correlation precisely, 
is more reliable than the previous mean-field-type calculations. 
We have compared the total energy for various pairing symmetries 
in a small-$J/|t|$ region, changing the doping rate $\delta$ 
and the sign of $t$. 
We consider $s$, extended-$s$, $d$, and $d$+$id$ waves for singlet 
pairing, and $p$+$ip$ and $f$ waves for triplet pairing. 
Furthermore, we have studied the stability of Nagaoka ferromagnetism 
and inhomogeneous states. 
Main results are summarized as follows.
\par

(1) For $t>0$, the $d$+$id$-wave state, which breaks time reversal 
symmetry, is dominant for $0<\delta\lsim 0.24$ ($J/t=0.3$), but the 
simple $d$-wave state has a quite close energy to that of the $d$+$id$ 
wave for a similar range of $\delta$. 
The stability of the superconducting state is rapidly lost, as the 
carrier density goes away from half filling. 
For $\delta\sim 0$, phase separation may take place. 
See Fig.~\ref{fig:phased}(a). 
\par

(2) For $t<0$, in a wide parameter range of our concern ($0<\delta\lsim 0.96$, 
$0\le J/t\lsim 0.7$), the Nagaoka ferromagnetic state defeats the 
$d$+$id$-wave state, which is the most stable among the superconducting 
states studied. 
See Fig.~\ref{fig:phased}(b). 
\par

(3) Triplet pairing is not realized in the $t$-$J$ model, although 
an $f$-wave state becomes slightly more stable than the normal state 
in the special case of $\delta=0.5$ and $t<0$, where the Fermi surface 
overlaps the van Hove singularity points. 
\par

These results broadly support the previous mean-field-type 
studies,\cite{Kumar,Baskaran,Wang,Ogata} and is consistent with the 
high-temperature expansion studies.\cite{Koretsune} 
Now, let us check these results for the triangular $t$-$J$ model 
in the light of a model of Na$_{0.35}$CoO$_2\cdot$1.3H$_2$O. 
Within the $t$-$J$ model with small $J/|t|$, the range of steady 
superconductivity is limited to $0<\delta\lsim 0.24$ with $t>0$, and 
the pairing symmetry is the $d$+$id$ (or possibly pure $d$) wave. 
In experiment, although the pairing symmetry of 
Na$_{0.35}$CoO$_2\cdot$1.3H$_2$O has not yet established, 
it is likely that its gap has a line node,\cite{Fujimoto,Ishida} 
and does not break the time-reversal symmetry.\cite{Higemoto,Uemura} 
Furthermore, the carrier density at which superconductivity appears 
is limited to a narrow range near $\delta=0.3$.\cite{Cava} 
Therefore, it is difficult to conclude that the $t$-$J$ model 
successfully describes the properties of Na$_{0.35}$CoO$_2\cdot$1.3H$_2$O 
in the present experimental situation. 
Provided that this discordance originates in the theoretical side, 
we discuss some related issues briefly in the remainder. 
\par

As mentioned in \S1, the results of triplet, especially $f$-wave, 
pairing were obtained by applying weak-coupling approaches to 
Hubbard-type models.\cite{Ikeda,Y.Tanaka} 
Among them, Kuroki \etal\ paid attention to the role of $e_g'$ 
band\cite{Kuroki}, which broadly corresponds to the $t<0$ band. 
They showed using a FLEX approximation that the $f$-wave state is 
the most stable. 
Thus, it is interesting to confirm whether the stabilization of 
the $f$ wave found for $\delta\sim 0.5$ and $t<0$ in the $t$-$J$ 
model is connected to this weak-coupling $f$ wave in the Hubbard model. 
Actually, the Hubbard-type model became tractable by a VMC method 
due to a recent improvement of the variational wave functions.\cite{YTOH}
\par

In this paper, we have used a single-band model as a first step 
to study the Co-based oxide. 
However, there exists a possibility that the multiband effect plays 
an essential role for superconductivity.\cite{Koshibae,Yata} 
Actually, recent FLEX and perturbation calculations \cite{Mochizuki} 
performed on multi-band Hubbard models yielded results of dominant 
$f$ and $p$ waves. 
Here, the $e_g'$ band again makes a key contribution to superconductivity. 
To study this effect for large $U/t$ is also an interesting 
remaining issue. 
\par

In the connection of triplet pairing, we finally remark that the pairing 
symmetry of Na$_{0.35}$CoO$_2\cdot$1.3H$_2$O can be determined by the 
phase sensitive experiments like tunneling effect and Josephson effect 
through the Andreev bound states. 
\cite{Asano,Y.Tanaka2,Kashiwaya,Y.Tanaka3,Y.Tanaka5}
In particular, to identify the triplet pairing, it is effective to use 
the anomalous proximity effect predicted very recently.\cite{Y.Tanaka4}
\par

%-----------------------------------------------------------------------

\begin{acknowledgments}
The authors thank Kazuhiko Kuroki for useful discussions. 
This work is partly supported by Grant-in-Aids from the Ministry of 
Education, Culture, Sports, Science and Technology, 
by the Supercomputer Center, ISSP, University of Tokyo, 
and by NAREGI Nanoscience Project, Ministry of Education, Culture, Sports, 
Science and Technology, Japan, which enables us to carry out the calculations 
on the computers at the Research Center for Computational Science, Okazaki 
National Research Institutes. 
\end{acknowledgments}

%%%%%%%%%%%%%%%%%%%%%%%%%%%%%%%%%%%%%%%%%%%%%%%%%%%%%%%%%%%%%%%%%%%%%%%%%

%%%%%%%%%%%%%%%%%%%%%%%%%%%%%%%%%%%%%%%%%%%%%%%%%%%%%%%%%%%%%%%%%%%%%%%%%


\begin{thebibliography}{99}
\bibitem{note:Eg}
As usual, we use the notation $e'_g$ to distinguish it from $e_g$.
\bibitem{note:ext.s} For example, for $\delta=0.22$, the extended-$s$ 
wave (1) has a minimum at a finite value of $\Delta$ for $J/t>0.5$ 
for $t>0$, and for $J/t>0.8$ for $t<0$. 
For the type (2), the energy minimum is at $\Delta=0$ for at least 
$J/t<1.0$. 
However, we leave it for a future study whether the extended-$s$ 
wave becomes stable for larger values of $J/|t|$, as in the square 
lattice.\cite{Dagotto,YO}
\bibitem{half-filling} At half filling, the pure $d$ wave has a lower
value of $E_{\rm tot}$ than the $d$+$id$ wave, for finite systems. 
Checking the system-size dependence ($L=4,6,12$ and 18), however, 
we have found the values of the two symmetries seem degenerate 
in the $L\rightarrow\infty$ limit. 
In reality, an antiferromagnetism with $120^\circ$ structure, which 
is outside the scope of this study, is realized at $\delta=0$. 
We believe this N\'eel structure immediately breaks as carriers are 
doped, due to the substantial frustration.
\bibitem{note:degeneracy} In this condition, we replace $E(0)$ with 
$E(\Delta\rightarrow 0_+)$, because the two values do not coincide.
\bibitem{note:hf} Within the present SC wave functions, the best 
value of $E_J/J$ ($d$+$id$ wave, $L=18$) at half filling is 
$-1.2369\pm 0.0004$, which is approximately 5\% higher than a 
value obtained using a projector Monte Carlo method, 
$-1.2958\pm 0.0001$. \cite{Capriotti} 
Thus, $E^{\rm PS}<E^{d+id}$ holds for $\delta\rightarrow 0$, because, 
then, $E_t$ vanishes.
\end{thebibliography}
\end{document}